\documentclass[journals]{IEEEtran}
\IEEEoverridecommandlockouts
\usepackage{cite}
\usepackage{url}
\usepackage{amsmath,amssymb,amsfonts}
\usepackage{algorithmic}
\usepackage{graphicx}
\usepackage{textcomp}
\usepackage{xcolor}
\usepackage{multirow}
\usepackage{subcaption}
\usepackage{tikz}
\usepackage{fancyhdr}

\newcommand{\tikzcircle}[2][red,fill=red]{\tikz[baseline=-0.5ex]\draw[#1,radius=#2] (0,0) circle ;}%

\newcommand{\customfooterimage}{figures/berc_banner.png}

\fancypagestyle{firstpage}{
  \lfoot{\includegraphics[width=\textwidth,height=1.75cm]{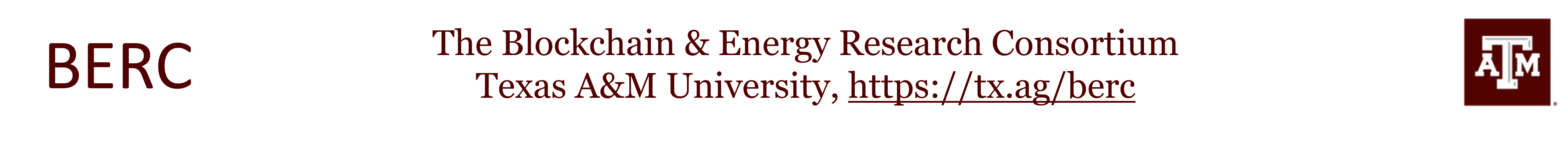}}
  \rfoot{}
}

\begin{document}

\title{Electromagnetic Transient Model of Cryptocurrency Mining Loads for Low-Voltage Ride Through Assessment in Transmission Grids}

\author{Anindita Samanta,
	Subir Majumder,~\IEEEmembership{Member,~IEEE,}
    Hasan Ibrahim,~\IEEEmembership{Student Member,~IEEE,}
	Prasad Enjeti,~\IEEEmembership{Fellow,~IEEE,}
    Le Xie,~\IEEEmembership{Fellow,~IEEE}
    \thanks{Anindita Samanta, Subir Majumder, Hasan Ibrahim, Prasad Enjeti, and Le Xie are with the Department of Electrical Engineering and Computer Sciences, Texas A\&M University, USA. (Corresponding author: le.xie@tamu.edu)}
	\thanks{This work is supported in part by Texas A\&M Energy Institute and in part by the Blockchain and Energy Research Consortium.} 
}

\maketitle
\thispagestyle{firstpage} 

\begin{abstract}
In this paper, we developed an Electromagnetic Transient (EMT) model tailored for large cryptocurrency mining loads to understand the cross-interaction of these loads with the electric grid. The load model has been built using Electromagnetic Transients Program (EMTP) software. We have cross-validated the performance of the EMT model of the load with commercial application-specific integrated circuit miners, typically used by large-scale mining facilities, by comparing their low-voltage ride-through (LVRT) capabilities. Subsequently, LVRT capabilities of the large-scale miners have been tested against various fault scenarios both within the miner's remote facility as well as at one of the distant buses of the interconnected grid. The significance of this model lies in its scalability to accommodate larger blocks of mining loads and its seamless integration into a larger electric grid.
\end{abstract}

\begin{IEEEkeywords}
Cryptocurrency mining loads, Electromagnetic model, EMTP, Low Voltage Ride Through, Transient studies
\end{IEEEkeywords}

\section{Introduction}

\begin{figure}
\centering
\includegraphics[width=0.40\textwidth]{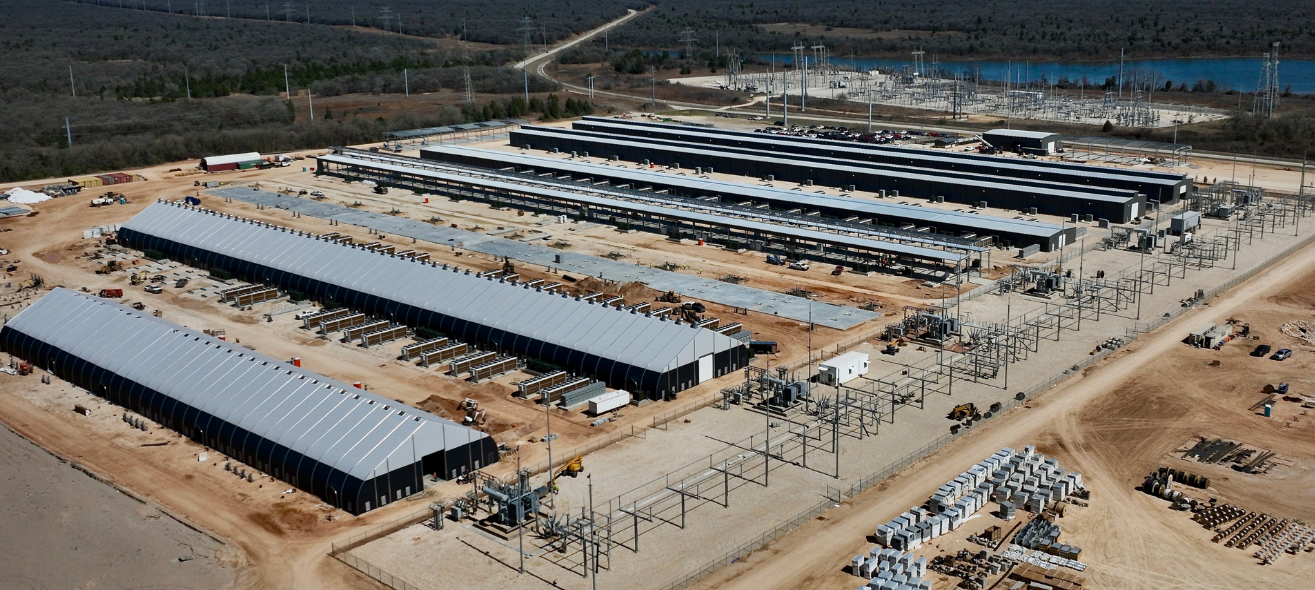}
 \caption{Mining firm  Riot Platforms, Inc. with own substation of 750 MW located in Rockdale, Texas \cite{mining-illustration}.}
 \label{fig:facility}
 \end{figure}

\IEEEPARstart{O}{ver} the past decade, blessed with abundant wind and solar resources, the Texas power grid has witnessed a significant increase in renewable energy penetration, accounting for almost 40\% of the state's total generation capacity \cite{ercot_gen_que}. This was partly the reason that drew attention from many emerging electric loads, such as large cryptocurrency mining operations \cite{forbes}. These crypto-mining industrial facilities with their large sitewide energy demand (a typical facility has a load of $\sim$ 75MW, see Fig. \ref{fig:facility} that shows a typical mining firm), integrated with power-electronic converters, play a crucial role in providing frequency support to the electricity grid based on price signals. However, these resources are similar to wind and solar resources, and operate on inverter-based (IBR) systems with distinct dynamics in contrast with  conventional synchronous generators and loads. Notably, recent outage events at wind and solar farms integrated with IBR-based resources in the Panhandle and Odessa regions of Texas have raised significant concerns \cite{ercot_Odessa},\cite{ercot_West_Texas}. Such incidents were considered to be associated with a lack of proper electromagnetic transient responses from IBR-based resources \cite{EMT-Req}, which would likely exacerbate with increasing penetration of crypto-mining loads. 


Owing to ERCOT's (major power grid operator operating within the state of Texas, USA) interconnection guidelines \cite{ercot_IBR_norm}, IBR-based generating resources should maintain connectivity even during periods of low grid voltage -- a requirement known as Low Voltage Ride Through (LVRT) capability. While ERCOT's existing framework does not impose specific mandates on these IBR-based large loads, recent incidents in West Texas on October 12, 2022 \cite{ercot_West_Texas} vividly demonstrated the repercussions of a low voltage event, triggering four cascading faults and ultimately leading to an outage exceeding 400MW, which included large-flexible loads, such as crypto-mining facilities. The recorded minimum voltage during this episode plummeted to 0.36 \textit{pu}. 
With a surge in requests for interconnection from these cryptocurrency loads \cite{ercot_load_que}, ERCOT has been scrambling to conduct power grid dynamic performance analysis for developing newer LVRT standards for these loads.

In regards to the modeling efforts for crypto-mining facilities, efforts have been made toward the use of power quality analyzers within a farm. In this context, Wheeler et al. have conducted power factor and harmonic analysis for a physical site with S9 AntMiners as processing units \cite{wheeler2018power}. As demonstrated in this study, which is also corroborated by our lab tests (see Fig. \ref{fig:startup}) available in \cite{IEEEECCE2023}, the voltage waveform does not suffer from distortions owing to being connected to an ideal power source, but, the current waveform shows distortion during startup. In \cite{IEEEECCE2023}, we also conducted power quality analysis at two industrial facilities where we observed similar performance. During the steady state, the power factor remains between 0.994 and 0.995 leading. However, these loads show non-linear characteristics during power system transients. While data center loads may not be as flexible as crypto-mining loads, being majorly computing demand, they are expected to behave as a crypto-mining load during transients. Regarding this, \cite{harinath2016critical} analyses the utilization of automatic power factor corrector to compensate for the required reactive power demand for IT loads and UPSs emulating datacenter. Ref. \cite{9803460} also showed high harmonics content in the power supply measurements, significantly distinct from cryptocurrency mining loads.



The need for accurate modeling of crypto-mining facilities through dynamic load modeling, frequency scan, and eigenvalue analysis has already been discussed in the existing literature \cite{10253366}. However, as discussed in \cite{10253366}, the electromagnetic transient (EMT) model of these loads for their transient performance analysis is largely missing. In a recent study, the LVRT capabilities of a S19 miner have been characterized in laboratory setup utilizing actual crypto-mining power supply \cite{IEEEECCE2023}. However, in order to set standards for grid interconnection in Texas, the development of an EMT model for these power supplies, appropriate analysis of their transient performance, and their interaction with other resources in the power grid are extremely important. The contribution of this proposed work is therefore twofold:


\begin{itemize}
    \item[i.] This paper aims to develop an EMT model of a cryptocurrency mining facility. Both, the time-domain response and voltage sag magnitude-duration capability characteristics of the developed model have been validated against a laboratory-scale crypto-mining load. The EMT model has been developed in the Electromagnetic Transients Program (EMTP) software. It has been made open source\footnote{Repository is available at: https://github.com/tamu-engineering-research/COVID-EMDA}.
    
    \item[ii.] The performance of the developed EMT model has been demonstrated utilizing a 120 kV 6-bus test system considering faults at two different locations: (i) at the crypto-miner's premises, and (ii) at bus 3 of the test system.

 
\end{itemize}


\section{Development of EMT Model of a Cryptocurrency Mining Load} \label{sec:2}

\begin{figure}
\centering
\includegraphics[width=0.40\textwidth]{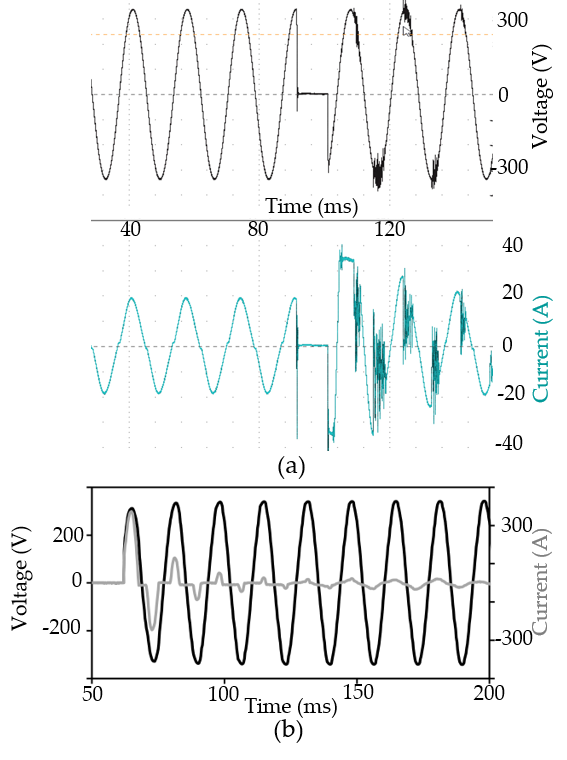}
 \caption{IV Characteristics of mining power supply: (a) Based on power supply available in the lab, (b) Startup time-domain voltage and current waveforms reproduced from \cite{wheeler2018power}.}
 \label{fig:startup}
 \end{figure}

Contrary to general-purpose computing resources, an industrial-scale cryptocurrency mining facility utilizes application-specific integrated circuit (ASIC) chips embedded in a hashing board. These hash boards are powered by the mining power supply. Typically these mining power supplies are driven by single-phase 240 V source, however, in recent times, the use of three-phase power supplies are also notable. In this analysis, we'll focus on the widely-used S19 Antminer with a 240 V single-phase power supply. Based on Fig. \ref{fig:startup}, and the power supply repair guide for the Antminer \cite{Antminer_powerguide}, we model the crypto-mining power supply as active power factor correction (PFC)-boost converters. Although diverse PFC converters, each with their unique dynamic performances, exist \cite{9887746}, we have opted for conventional boost PFC converters for their versatile uses.

\begin{figure}
\includegraphics[width=0.5\textwidth]{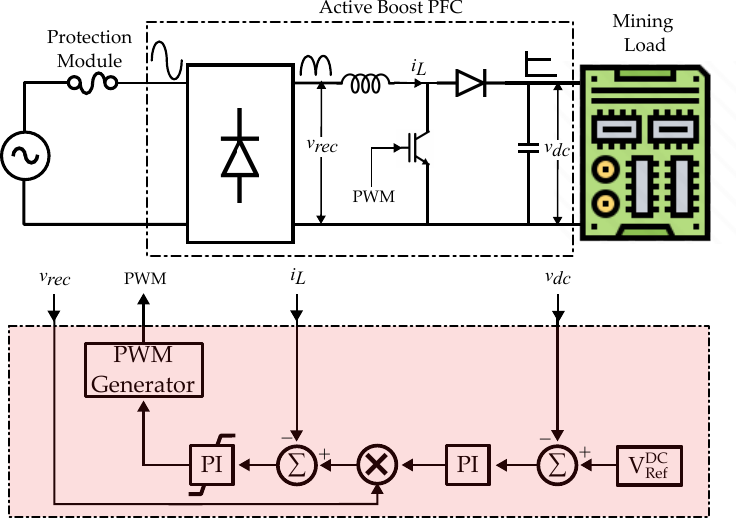}
 \caption{Block diagram of active PFC-Boost converter representing mining power supply.}
 \label{fig:blockdiag}
 \end{figure}

Electromagnetic Transient Program (EMTP) software is widely used for transient performance analysis \cite{IEEEexample:PM_book} of the power system. Here, we have developed switching-based models of an active PFC-boost converter in the EMTP to model the crypto-mining facility. A block diagram of this converter is provided in Fig. \ref{fig:blockdiag}.
The converter consists of a bridge rectifier at the input side and a boost circuit aiming to maintain a constant DC-bus voltage $v_{dc}$. As shown in the figure, the profile of rectified loads and the deviation in DC-bus voltage passed through the PI controller provide the references for the inductor current in the boost converter. The second PI controller with a limiter generates PWM signals for the IGBT within the converter to track the inductor current with the determined reference. The inductor and capacitors are designed to allow a certain amount of ripples in the voltage and current waveform. The gains of the PI controllers could be tuned for desired fastness in response from the converters. As described in \eqref{eq:1}, these loads behave as constant power loads (CPLs). Here, $V_{rms}(t)$, $I_{rms}(t)$ are the rms voltage and current magnitudes and $P_{av}$ is the average power drawn by the crypto-mining power supply.

\begin{equation}
    V_{rms}(t) I_{rms}(t) = P_{av} \implies \frac{\partial V_{rms}}{\partial t} I_{rms} +   V_{rms} \frac{\partial I_{rms}}{\partial t} = 0 \label{eq:1}
\end{equation}

\begin{figure}
\includegraphics[width=0.5\textwidth]{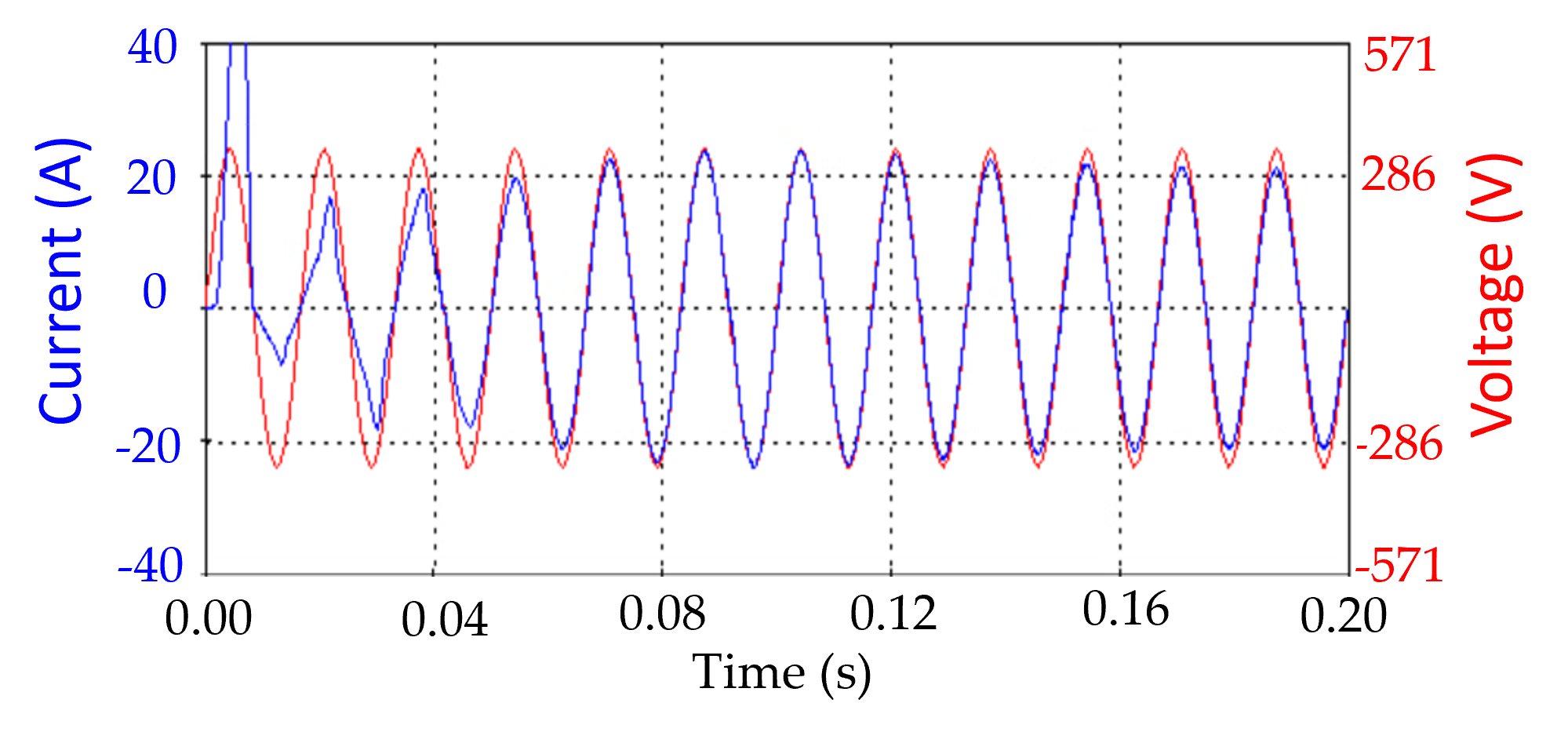}
 \caption{Start-up waveform for the EMT model of cryptocurrency miner power supply.}
 \label{fig:startupwaveformmodel}
 \end{figure}
 
\begin{figure}
\centering
\begin{subfigure}{0.45\textwidth}
   \includegraphics[width=1\linewidth]{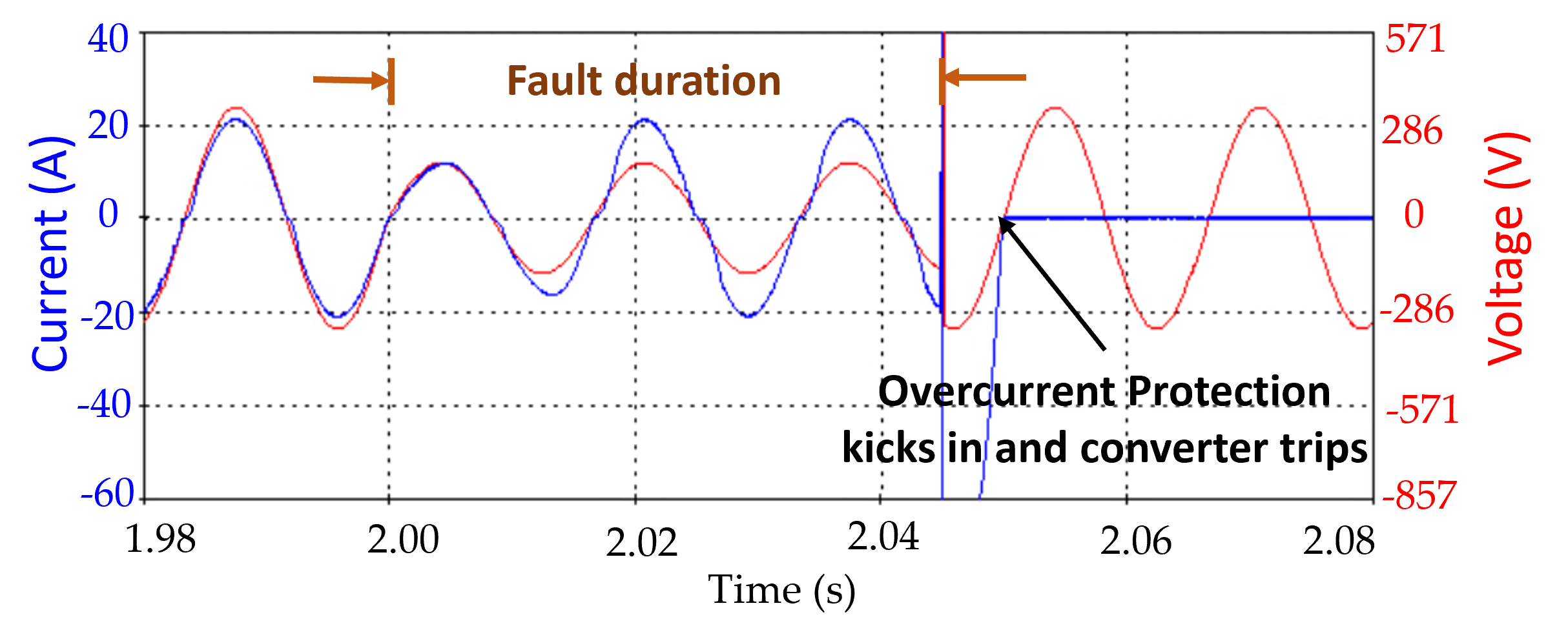}
   \caption{Simulated fault of 50\% pre-fault voltage for 45 ms duration.}
   \label{fig:Ng1} 
\end{subfigure}

\begin{subfigure}{0.45\textwidth}
   \includegraphics[width=1\linewidth]{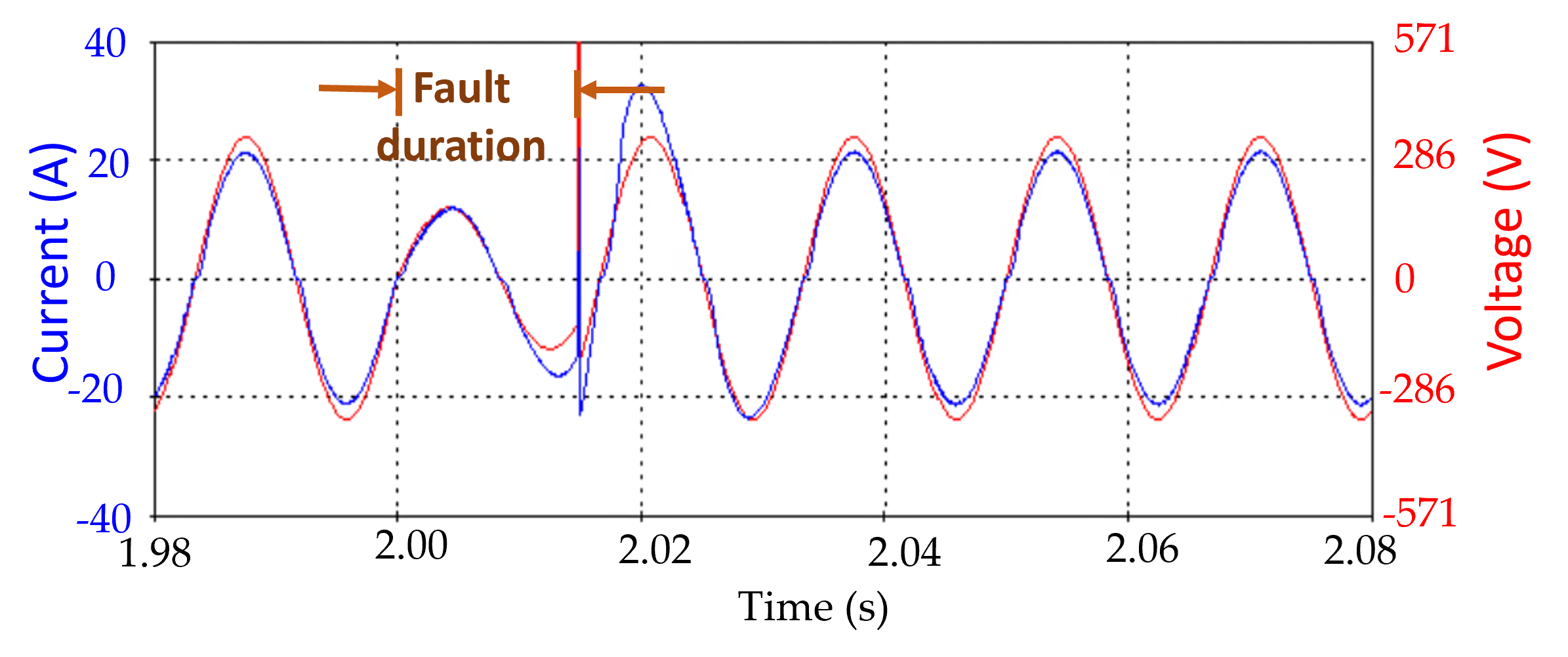}
   \caption{Simulated fault of 50\% pre-fault voltage for 16 ms duration.}
   \label{fig:Ng2}
\end{subfigure}

\caption{Time-domain response of the EMT model of the crypto-mining power supply.}
\label{fig:LVRT_EMT} 
\end{figure}

Being operated as a CPL implies that during the transients the product of the rms voltage and the current remains constant. The implications of this behavior are imminent across all the time-duration characteristics shown in this paper. During start-up, depending upon the DC-bus voltage, the converter would show inrush, which can be found in both Figs. \ref{fig:startup}(b) and \ref{fig:startupwaveformmodel}. Both of these figures also show that after the initial transient, the voltage and current become in phase. During sags, the converter current drops suddenly but grows as time progresses, as shown in Fig. \ref{fig:LVRT_EMT}. As the voltage recovers, the inrush kicks in as shown in Figs. \ref{fig:blockdiag}(a), and \ref{fig:LVRT_EMT}, which may lead to tripping of mining power supply. Fig. \ref{fig:blockdiag} demonstrates that during the zero-voltage condition, the current reference would be zero, and therefore, as shown in Fig. \ref{fig:startup}(a), the current drawn by the mining power supply would also be zero. Ideal voltage and current waveforms of the crypto-mining power supply could be given as follows:

\begin{equation}
    i(t) = \frac{P_{av}}{V_{rms}(t){}^2 + \epsilon} v(t)
\end{equation}

$\epsilon$ is a small positive real number prohibiting division by zero. $v(t)$ and $i(t)$ are instantaneous voltage and current drawn by the power supply.

\begin{figure}[t!]
\includegraphics[width=0.5\textwidth]{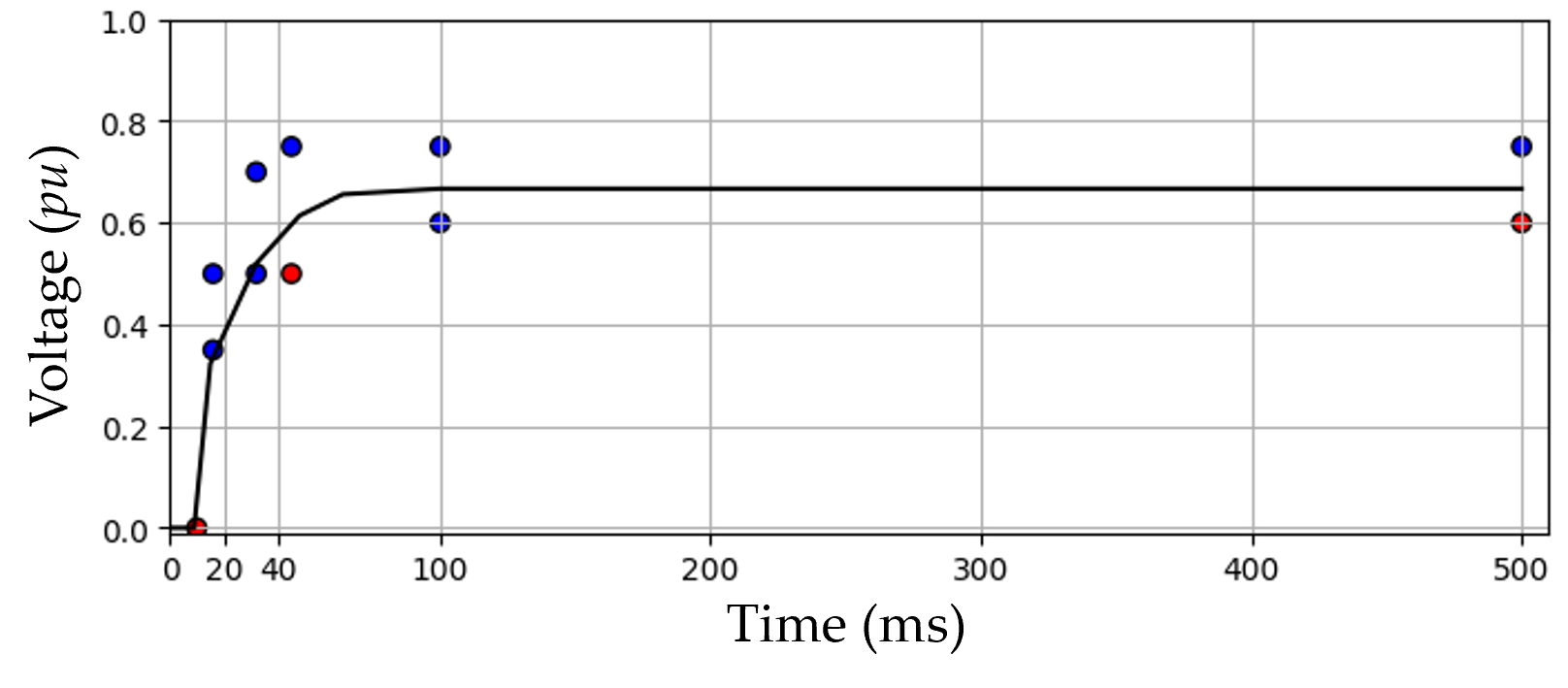}
 \caption{Validation with laboratory crypto-mining power supply. The solid black line corresponds to the laboratory results from the mining power supply. \tikzcircle[fill=red]{2pt} show experiments where the power supply was tripped. \tikzcircle[fill=blue]{2pt} show experiments where the power supply \underline{did not} trip. }
 \label{fig:mag-duraton}
 \end{figure}

The converters are designed to operate below certain loading capabilities, determining their overcurrent limits. They are also equipped with both over and under-voltage protection. During the inrush condition, if the overcurrent circuitry gets triggered, or the DC-bus voltage drops significantly, it may eventually cause the mining loads to trip. In regards to the protection circuitry of the model, we have used a simplified overcurrent protection, where, the crypto-mining power supply trips if the rms value of the current through the inductor exceeds a certain threshold. We have performed extensive tests with the developed model by connecting the developed model to an ideal power supply and subjecting it to voltage sags of various magnitudes and durations. For further model validation, we compared this sag magnitude-duration capability characteristics to a S19 ASIC miner available in our lab. The results are presented in Fig. \ref{fig:mag-duraton}. Individual measurements obtained out of the laboratory testing are provided in \cite{IEEEECCE2023}, and for representational simplicity, we have denoted it by a continuous line.

During model testing and simulation, it became evident that the severity of voltage sags and the current drawn during these events are influenced not only by the magnitude and duration of the sag but also by the specific point on the wave\cite{bollenUnderstandingPowerQuality2000}, (see case study section, where tripping of a partial set of converters is notable). Nevertheless, this additional information is absent in the voltage sag magnitude-duration characteristics. Hence, modeling the converter to align exclusively with magnitude-duration characteristics, while neglecting the broader converter protection system could lead to deviations from laboratory experiments. (see one of the test results deviates from the ideal scenario).

\section{Case Studies} \label{sec:3}

\begin{figure}[t!]
\includegraphics[width=0.47\textwidth]{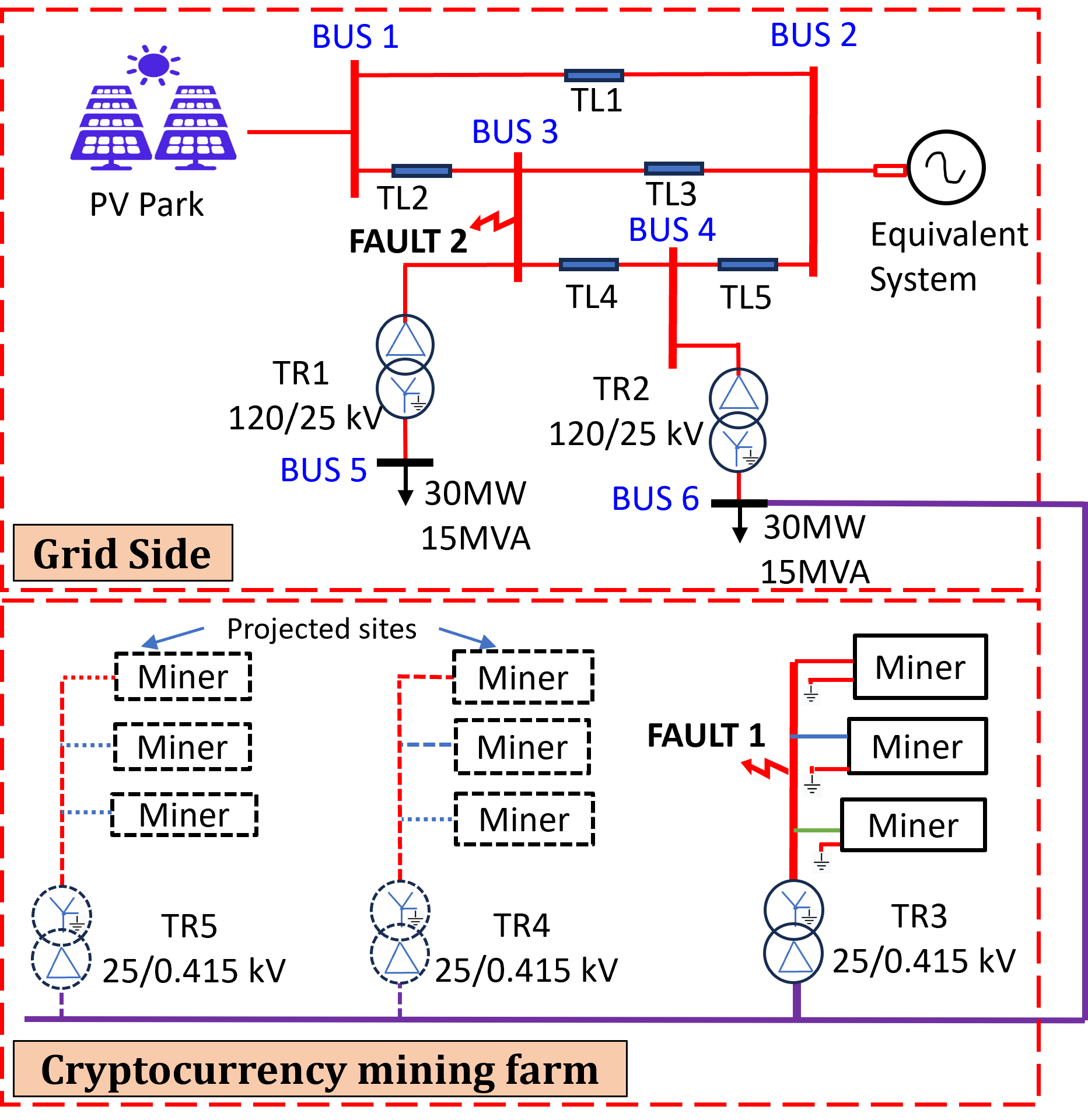}
 \caption{Cryptocurrency farm interconnected with 120 kV Test System. The performance of mining converters is compared for two distinct scenarios considering three-phase line-to-ground (LLLG) faults: (i) FAULT 1: fault within the cryptocurrency miner's premises, and (ii) FAULT 2: fault at bus 3.}
 \label{fig:farm}
 \end{figure}

The performance of the developed EMT model has been tested in a 120 kV, 60 Hz, 6 bus transmission network in EMTP. A detailed description of test system containing the solar PV park is provided in \cite{EMTP_PVPark}. A moderately sized crypto-mining facility of $\sim$ 1 MW with loads equally distributed across all three phases was considered. As shown in Fig. \ref{fig:farm}, the mining facility is connected with the rest of the transmission network through a $\Delta$ - Yg 25 kV/415 V transformer with a percentage impedance of 6 \% at Bus 6. Therefore, the miners are supplied with an input voltage at 240 V with the DC-bus voltage to be maintained at 400 V.

The non-mining loads in the test system are modeled as constant impedance loads and the transmission lines are modeled using constant parameter models. Both loads are 30 MW each and the solar PV park has a capacity of 75MVA. We consider the constant power output from the solar park across all scenarios and constant load demand from the crypto-mining facility at their respective rated capacity. The projected sites in Fig. \ref{fig:farm} represent the scalability of our model. The faults were initiated after the cryptocurrency miners reached a steady state. We explored multiple cases under each of these scenarios based on fault duration and fault impedance dictating the sag voltage magnitude.

\subsection{Fault applied at Load End}

\begin{figure}
\centering
\begin{subfigure}{0.45\textwidth}
   \includegraphics[width=1\linewidth]{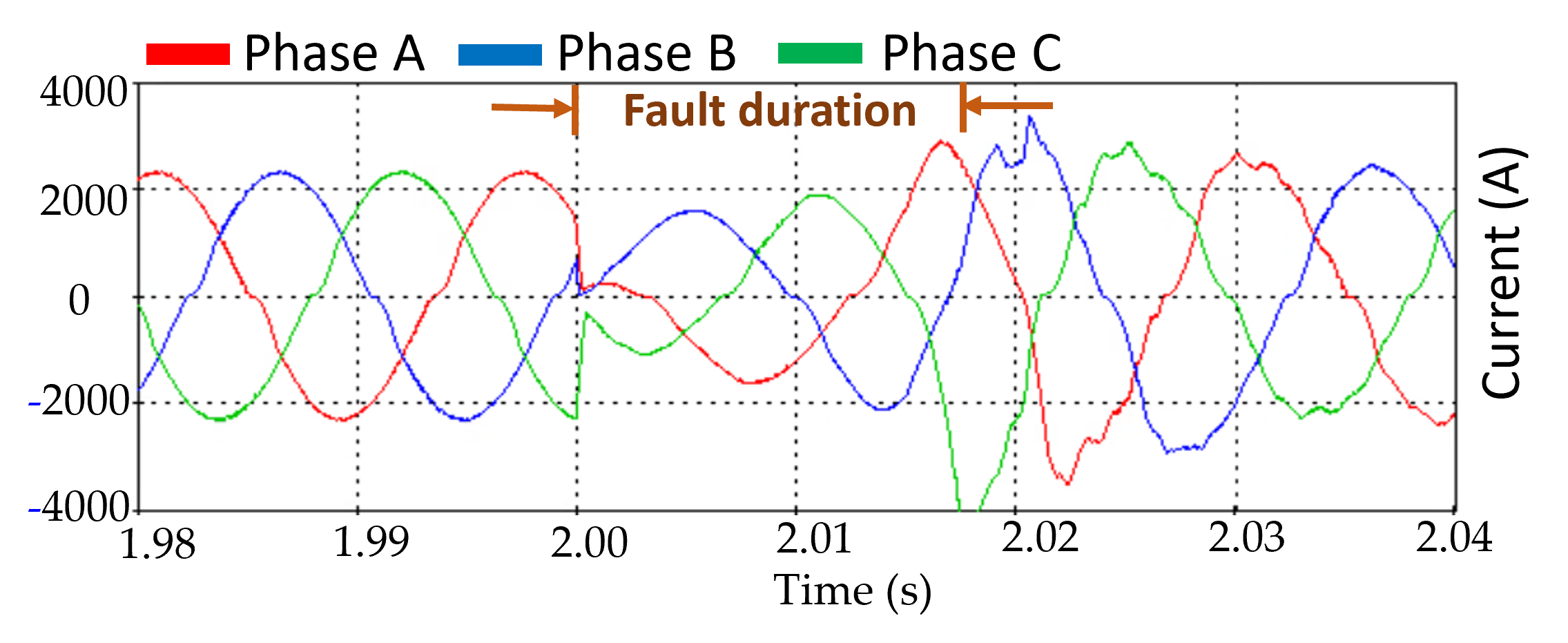}
   \caption{Current Waveforms of Miners (Aggregated) Facility.}
   \label{fig:Ng1} 
\end{subfigure}

\begin{subfigure}{0.45\textwidth}
   \includegraphics[width=1\linewidth]{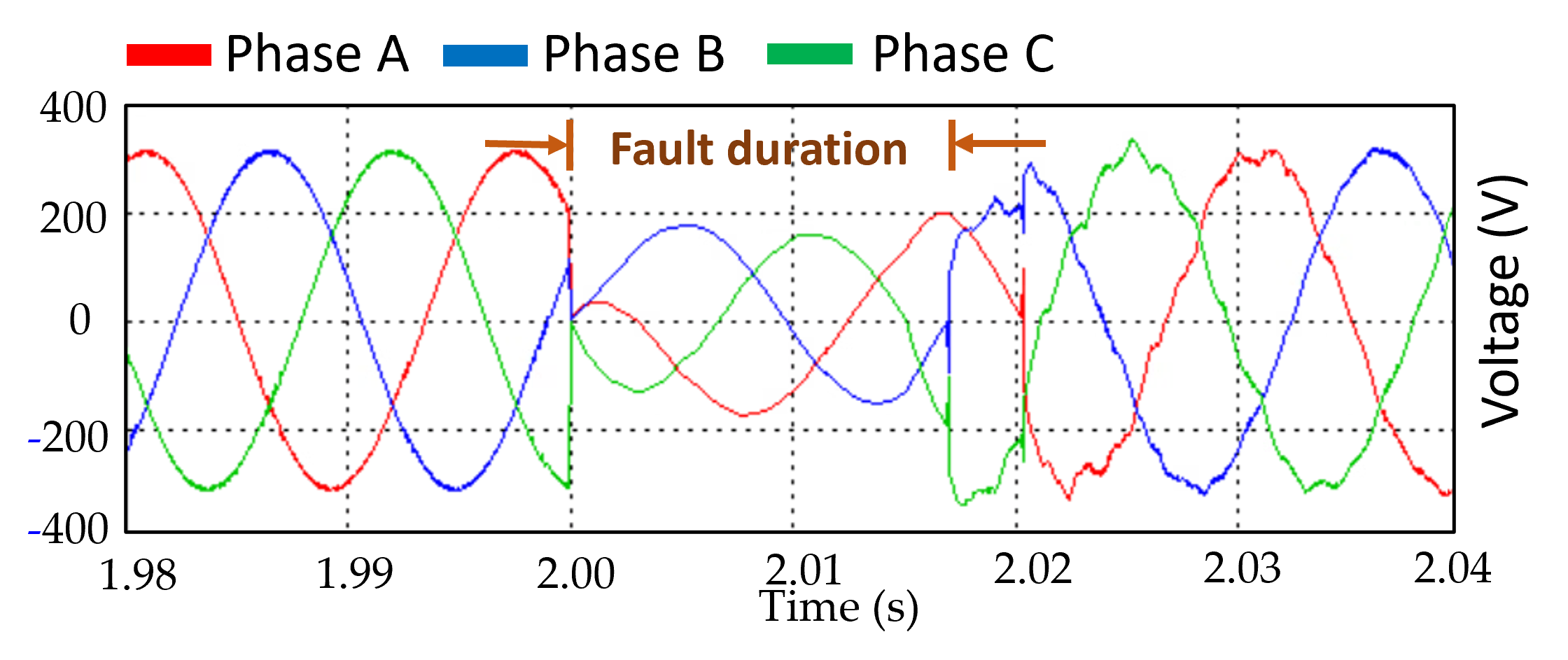}
   \caption{Fault Voltage at Miner Premises.}
   \label{fig:Ng2}
\end{subfigure}

\caption{Voltage and current at miner's premises with a 3-$\phi$ fault of 50\% pre-fault voltage of 15 ms duration applied at load end.}
\label{fig:Case2B} 
\end{figure}

\begin{figure}
\centering
\begin{subfigure}{0.45\textwidth}
   \includegraphics[width=1\linewidth]{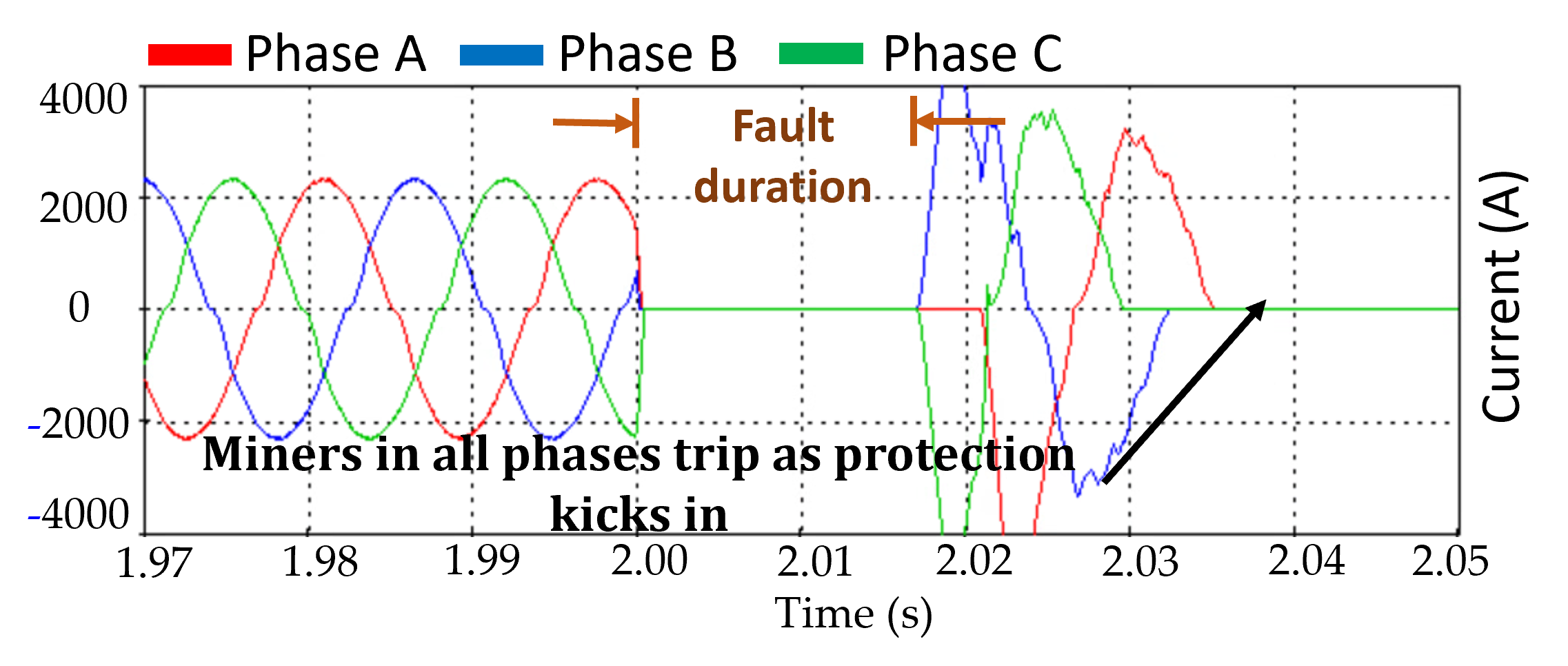}
   \caption{Current Waveforms of Miners (Aggregated) Facility.}
   \label{fig:Ng1} 
\end{subfigure}

\begin{subfigure}{0.45\textwidth}
   \includegraphics[width=1\linewidth]{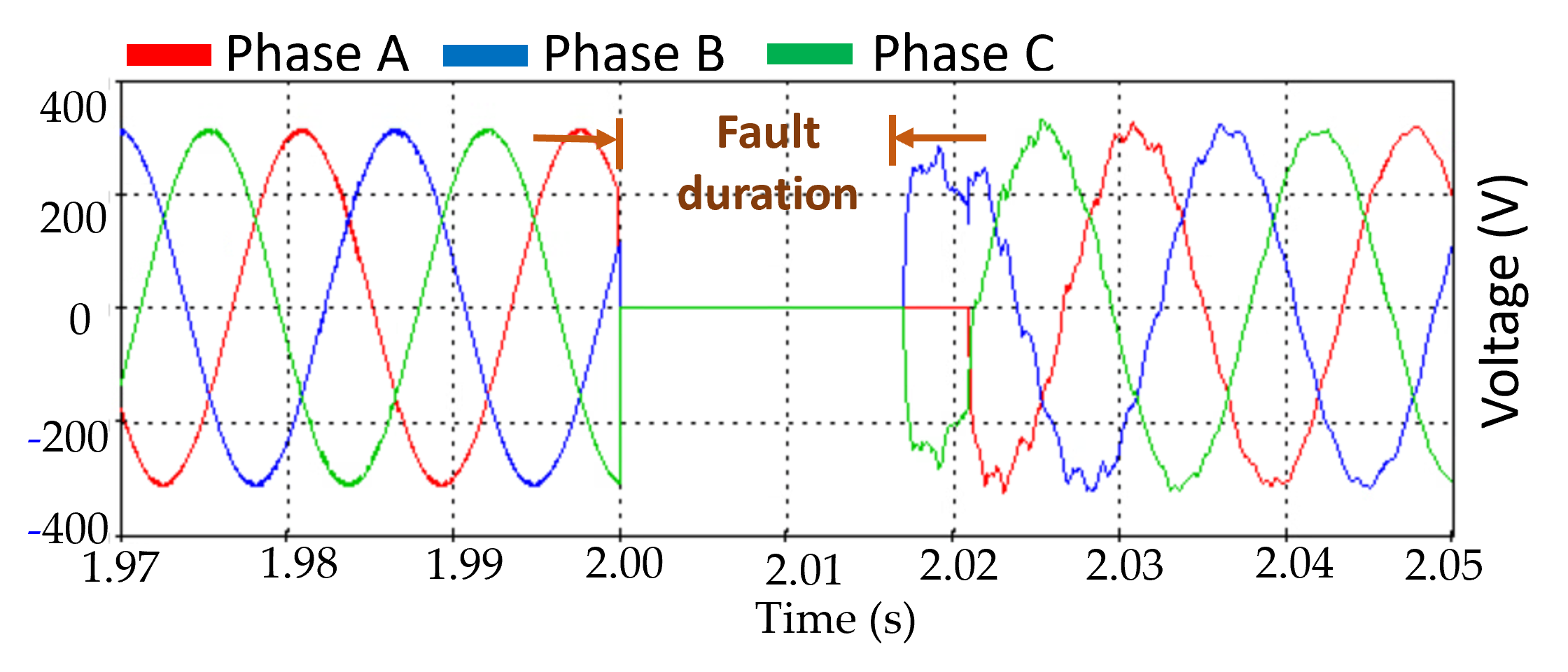}
   \caption{Fault Voltage at Miner Premises.}
   \label{fig:Ng2}
\end{subfigure}

\caption{Voltage and current at miner's premises with a bolted 3-$\phi$ fault of 15 ms duration applied at load end.}
\label{fig:Case4B} 
\end{figure}

Compared to Fig. \ref{fig:LVRT_EMT}, limited short-circuit capability implies that the post-fault network voltage may not remain as ideal sinusoid as depicted in Figs. \ref{fig:Case2B}(b) and \ref{fig:Case4B}(b). Here, we have compared the LVRT capability of the crypto-miners' power supply with a 3-$\phi$ bolted fault and a fault with 50\% pre-fault voltage, each for a duration of 15 ms. While the miners in all three phases were able to ride through with a fault of 50\% pre-fault voltage, none of the miners were able to ride through with the bolted fault, which is in line with the ride-through characteristics shown in Fig. \ref{fig:mag-duraton}. Also, it can be seen that after the miners were tripped, in Fig. \ref{fig:Case4B}(b), the voltage at the miners' premises contains a large amount of harmonics that tend to die out with time. It could safely be attributed to the PV farm.

\begin{table}[!ht]
\caption{LVRT Capability with Fault within Miner's Premises ($* \rightarrow$ all miners in one out of three phases trip). } \label{scenario1}
\resizebox{\columnwidth}{!}{\begin{tabular}{|c|cccc|}
\hline
\textbf{Sag Voltage} & \multicolumn{4}{c|}{\textbf{Duration of fault}}                                                                                                  \\ \cline{2-5} 
\textbf{(\% of Prefault)}                                      & \multicolumn{1}{c|}{\textbf{09ms}}   & \multicolumn{1}{c|}{\textbf{1 cycle (15ms)}} & \multicolumn{1}{c|}{\textbf{3 cycles (45ms)}} & \textbf{100ms}  \\ \hline
\textbf{75\%}                    & \multicolumn{1}{c|}{NO}              & \multicolumn{1}{c|}{NO}                      & \multicolumn{1}{c|}{NO}                       & NO              \\ \hline
\textbf{50\%}                    & \multicolumn{1}{c|}{NO}              & \multicolumn{1}{c|}{NO}                      & \multicolumn{1}{c|}{1 / 3 TRIP${}^*$}               & YES             \\ \hline
\textbf{25\%}                    & \multicolumn{1}{c|}{YES}             & \multicolumn{1}{c|}{YES}                     & \multicolumn{1}{c|}{YES}                      & YES             \\ \hline
\textbf{0\%}                     & \multicolumn{1}{c|}{YES}             & \multicolumn{1}{c|}{YES}                     & \multicolumn{1}{c|}{YES}                      & YES             \\ \hline
\end{tabular}}
\end{table}

We have conducted additional experiments to understand the LVRT capability of the mining facility if the faults are within the miner's premises, and the results are tabulated in Table \ref{scenario1}. As highlighted in Fig. \ref{fig:mag-duraton}, demonstrating the LVRT capability of crypto miner's power supply, with sag voltage of 75\% pre-fault voltage magnitude none of the miners should trip, and it has been validated in Table \ref{scenario1}. As with 50 \% pre-fault voltage and fault duration of 45 ms, all the miners in one of the phases out of all the three phases trips.
This can be attributed to the large post-fault inrush current drawn from the PCC which is a function of the point-on-wave. For a fault duration of less than 100 ms and with a sag voltage of less than 25 \% pre-fault voltage, none of the mining power supplies are able to ride through.

\begin{figure}
\centering
\begin{subfigure}{0.45\textwidth}
   \includegraphics[width=1\linewidth]{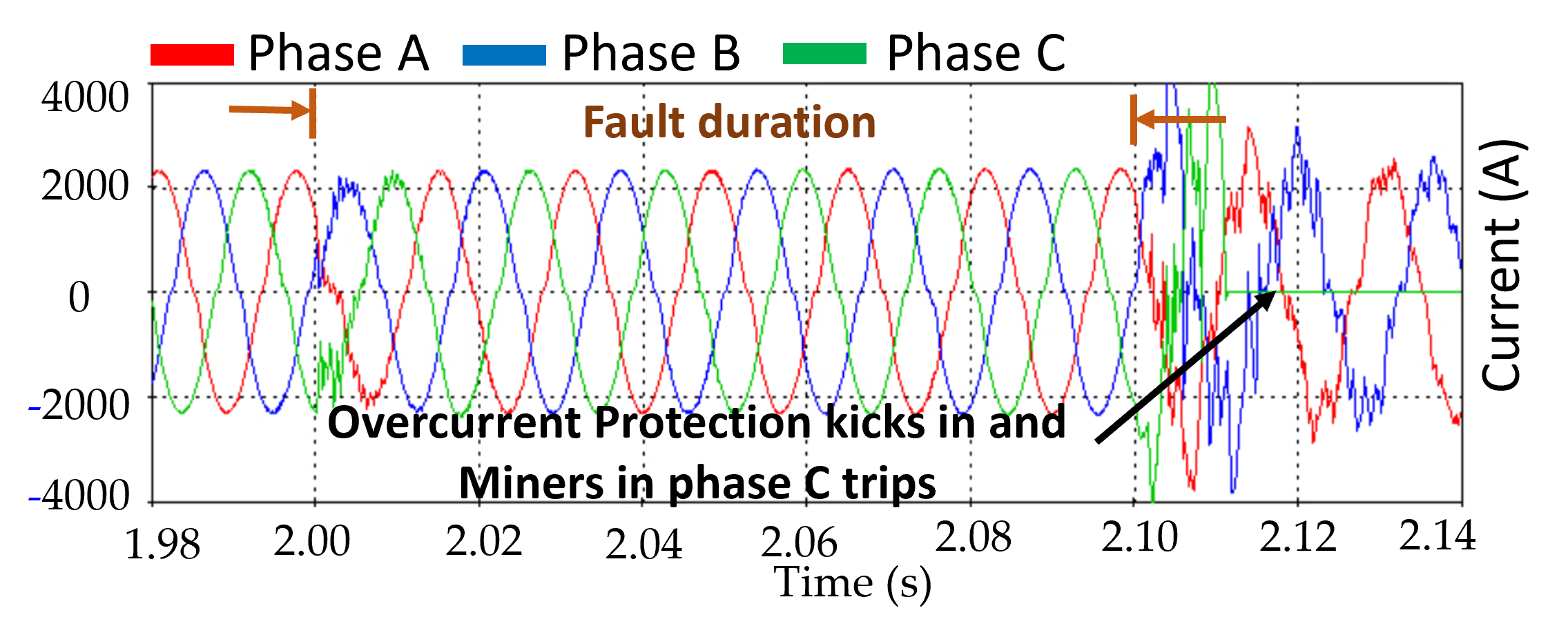}
   \caption{Current Waveforms of Miners (Aggregated) Facility.}
   \label{fig:Ng1} 
\end{subfigure}

\begin{subfigure}{0.45\textwidth}
   \includegraphics[width=1\linewidth]{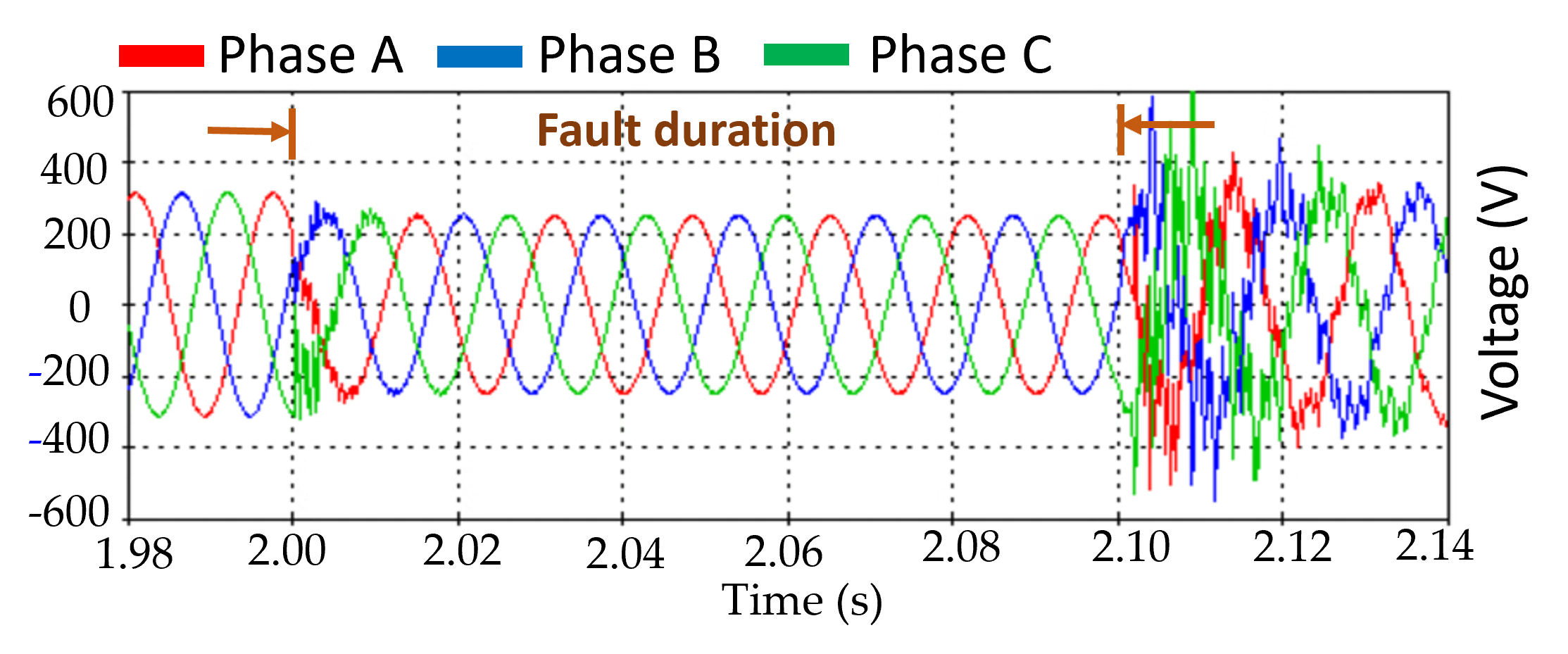}
   \caption{Voltage Waveforms at Miner Premises.}
   \label{fig:Ng2}
\end{subfigure}

\begin{subfigure}{0.45\textwidth}
   \includegraphics[width=1\linewidth]{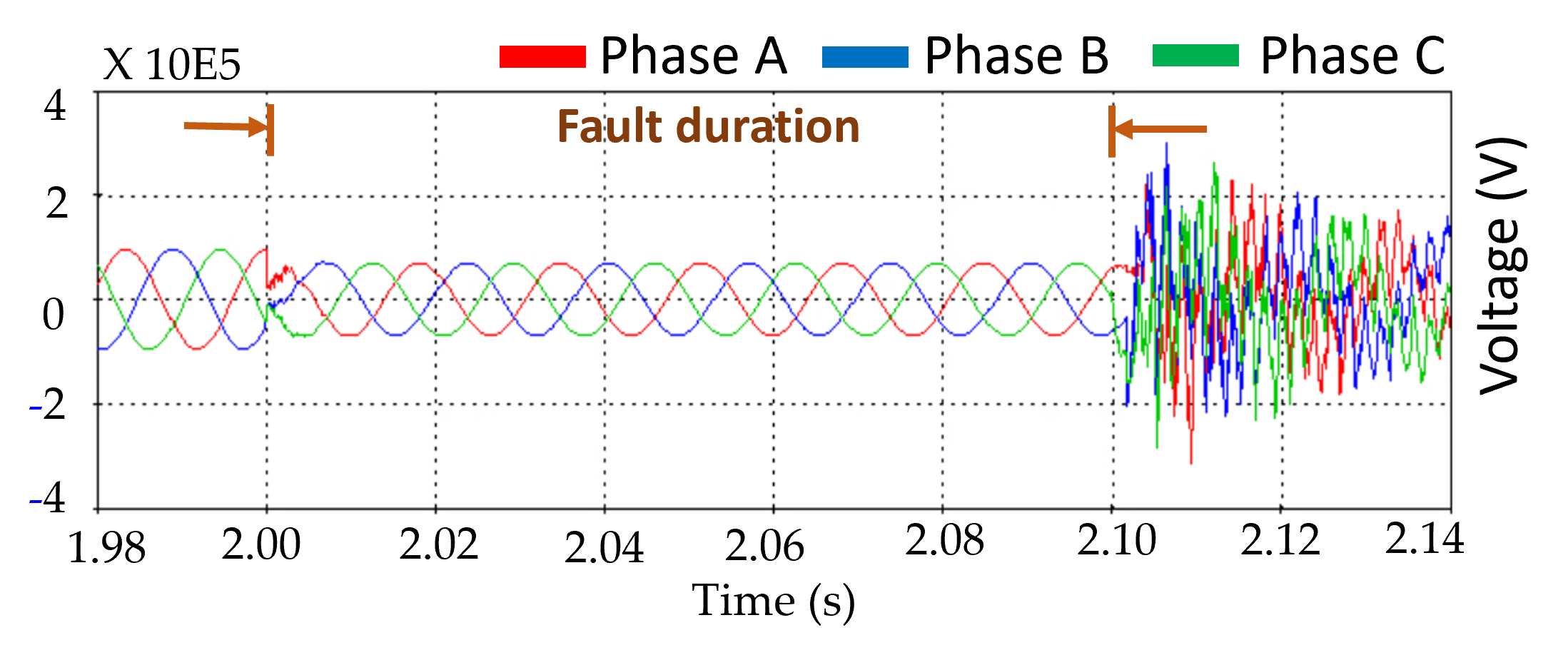}
   \caption{Fault Voltage at Bus 3.}
   \label{fig:Ng3} 
\end{subfigure}

\caption{Voltage and current at miner's premises and voltage at the faulted bus with a 3-$\phi$ fault of 75\% pre-fault voltage for a period of 100 ms duration applied at bus 3.}
\label{fig:S2_Case 1D} 
\end{figure}





\subsection{Fault at bus 3}

Owing to the transmission line's impedance, the sag voltage observed at the crypto-miners's premises would be less severe than in the previous scenario. However, as demonstrated in Fig. \ref{fig:S2_Case 1D}, the occurrence of faults near the PV park appears to inject a significant amount of harmonics.
This interacts with the controller of the mining power supply, and depending upon the phase it may lead to an outage of all the mining power supplies. We have further conducted an extensive number of experiments to understand the mining power supplies' LVRT capability. It can be seen that even with 75\% pre-fault voltage at the faulted node, or, a better voltage within the miner's premises, all the miners are not able to ride through. While not shown for brevity, it has also been observed that for a larger sag duration, more and more miners are unable to ride through even with 75\% pre-fault voltage. 

\begin{table}[!ht]
\caption{Miner's LVRT Capability with Fault at bus 3 ($* \rightarrow$ all miners in one out of three phases trip, $\dagger \rightarrow$ all miners in two out of three phases trip).}
\resizebox{\columnwidth}{!}{\begin{tabular}{|c|cccc|}
\hline
{\textbf{Sag Voltage}} & \multicolumn{4}{c|}{\textbf{Time duration of fault}}                                                                                                \\ \cline{2-5} 
\textbf{(\% of Prefault)}              & \multicolumn{1}{c|}{\textbf{09ms}}   & \multicolumn{1}{c|}{\textbf{1 cycle (15ms)}} & \multicolumn{1}{c|}{\textbf{3 cycles (45ms)}} & \multicolumn{1}{c|}{\textbf{100ms}}    \\ \hline
\textbf{75\%}                    & \multicolumn{1}{c|}{1 /3 TRIP${}^*$}       & \multicolumn{1}{c|}{1 /3 TRIP${}^*$}               & \multicolumn{1}{c|}{1 /3 TRIP${}^*$}                & \multicolumn{1}{c|}{1 /3 TRIP ${}^*$}                \\ \hline
\textbf{50\%}                    & \multicolumn{1}{c|}{1 /3 TRIP${}^*$}       & \multicolumn{1}{c|}{2 /3 TRIP${}^{\dagger}$}               & \multicolumn{1}{c|}{2 /3 TRIP${}^{\dagger}$}                & \multicolumn{1}{c|}{YES}                     \\ \hline
\textbf{25\%}                    & \multicolumn{1}{c|}{1 /3 TRIP${}^*$}       & \multicolumn{1}{c|}{YES}                     & \multicolumn{1}{c|}{YES}                      & \multicolumn{1}{c|}{YES}               \\ \hline
\textbf{0\%}                     & \multicolumn{1}{c|}{2 /3 TRIP${}^{\dagger}$}       & \multicolumn{1}{c|}{YES}                     & \multicolumn{1}{c|}{YES}                      & \multicolumn{1}{c|}{YES}              \\ \hline
\end{tabular}}
\end{table}

\section{Conclusion} \label{sec:4}

This paper provides a detailed switching-based EMT model of a typical cryptocurrency mining facility based on an active power factor correction (PFC) boost converter first by validating with a laboratory-based crypto-mining power supply, and then by integrating and testing it with a small-scale transmission system with a PV farm. The efficacy of the mining facility in terms of its LVRT capability was compared considering two fault scenarios: (i) fault within the miner's premises, and (ii) fault near the PV farm. The resulting harmonics from the PV farm in the aftermath of the fault can interact with the controllers of the cryptocurrency miner's power supply resulting in an outage of these cryptominers, which would be challenging, especially in a network with a low short circuit ratio. While the developed model of the cryptocurrency miners is highly scalable, the use of a detailed switching model hinders large-scale performance analysis of multiple cryptocurrency mining facilities in a larger system. Therefore, in the future, we would like to develop an average value model and utilize it for larger-scale performance analysis.

\bibliographystyle{IEEEtran}
\bibliography{ref.bib}

\begin{thebibliography}{10}
\providecommand{\url}[1]{#1}
\csname url@samestyle\endcsname
\providecommand{\newblock}{\relax}
\providecommand{\bibinfo}[2]{#2}
\providecommand{\BIBentrySTDinterwordspacing}{\spaceskip=0pt\relax}
\providecommand{\BIBentryALTinterwordstretchfactor}{4}
\providecommand{\BIBentryALTinterwordspacing}{\spaceskip=\fontdimen2\font plus
\BIBentryALTinterwordstretchfactor\fontdimen3\font minus \fontdimen4\font\relax}
\providecommand{\BIBforeignlanguage}[2]{{%
\expandafter\ifx\csname l@#1\endcsname\relax
\typeout{** WARNING: IEEEtran.bst: No hyphenation pattern has been}%
\typeout{** loaded for the language `#1'. Using the pattern for}%
\typeout{** the default language instead.}%
\else
\language=\csname l@#1\endcsname
\fi
#2}}
\providecommand{\BIBdecl}{\relax}
\BIBdecl

\bibitem{ercot_gen_que}
{Pablo Vegas}, ``Ercot public presentation: Item 5 : Ceo update - revised,'' Published in 31, August 2023, [Online]. Available : ~\url{https://www.ercot.com/files/docs/2023/08/28/5%20CEO%20Update%20REVISED.pdf}.

\bibitem{forbes}
{Jason Brett}, ``Forbes: Texas poised to be a world leader in bitcoin and blockchain,'' Published in Forbes 02, October 2021, [Online]. Available:~\url{https://www.forbes.com/sites/jasonbrett/2021/10/02/texas-poised-to-be-a-world-leader-in-bitcoin-and-blockchain/?sh=6e2db00a715b}.

\bibitem{ercot_Odessa}
{Dan Woodfin}, ``Ercot public presentation: Item 7.2.1: Inverter based resource and large load ride through events: Background and mitigation,'' Published in 19, June 2023, [Online]. Available:~\url{www.ercot.com}.

\bibitem{ercot_West_Texas}
{Agee Springer, Evan Rowe, Evan Neel}, ``Ercot public presentation: Ibrtf - lfl interconnection \& resource adequacy,'' Published in 05, April 2023, Texas A\& M University.

\bibitem{EMT-Req}
{ERCOT IBRTF Meeting - NERC}, ``Reliability guidelines: Electromagnetic transient (emt) modeling for bps-connected inverter-based resources – requirements and verification practices \& emt task force,'' [Online] Available:~\url{https://www.ercot.com/files/docs/2023/01/20/RG - EMT Modeling and Simulation - EMT TF - ERCOT IBRTF 2023.pdf}.

\bibitem{mining-illustration}
``North america's largest bitcoin mining facility by developed capacity,'' [Online] Available:~\url{https://www.riotplatforms.com/bitcoin-mining/whinstone-u-s}.

\bibitem{ercot_IBR_norm}
{Stephen Solis}, ``Ercot public presentation: Ibrtf - nogrr245 - inverter-based resource (ibr) ride-through requirements,'' Published in 20, January 2023, [Online]. Available :~\url{https://www.ercot.com/files/docs/2023/01/20/NOGRR245%20IRR%20Ride%20Through%20Requirements_IBRTF_01202023.pptx}.

\bibitem{ercot_load_que}
``Ercot public presentation:large load interconnection status,'' Published in 17, February 2023, \ [Online]. Available : ~\url{https://www.ercot.com/files/docs/2023/02/17/LLI%20Queue%20Status%20Update%20-%202023-02-17.pdf}.

\bibitem{wheeler2018power}
K.~A. Wheeler, A.~W. Bowers, C.~H. Wong, J.~Y. Palmer, and X.~Wang, ``A power quality and load analysis of a cryptocurrency mine,'' in \emph{2018 IEEE Electrical Power and Energy Conference (EPEC)}.\hskip 1em plus 0.5em minus 0.4em\relax IEEE, 2018, pp. 1--6.

\bibitem{IEEEECCE2023}
S.~Almubarak, H.~Ibrahim, D.~Singhania, and P.~Enjeti, ``Energy consumption \& power quality in bitcoin mining facilities in texas,'' in \emph{2023 IEEE Energy Conversion Conference and Expo}, 2023, pp. 1--5.

\bibitem{harinath2016critical}
P.~Harinath, K.~Kamaleswaran, M.~Venkateshwaran, C.~Sreenath, S.~Prabhakaran, and V.~Kirubakaran, ``A critical analysis of power quality issues in data center,'' in \emph{2016 Biennial International Conference on Power and Energy Systems: Towards Sustainable Energy (PESTSE)}.\hskip 1em plus 0.5em minus 0.4em\relax IEEE, 2016, pp. 1--6.

\bibitem{9803460}
K.~R. Raguž, M.~Miletić, V.~Zeleničić, D.~Sumina, I.~Erceg, and Z.~Nastasić, ``Analysis of current harmonics in data center power system,'' in \emph{2022 45th Jubilee International Convention on Information, Communication and Electronic Technology (MIPRO)}, 2022, pp. 153--157.

\bibitem{10253366}
S.~Mohan, S.~Maleki, M.~Shirinzad, B.~Yancey, H.~Trahan, and R.~Ayass, ``Load modeling impact on system stability and guidelines for stability studies on an islanded system with grid-forming inverters,'' 2023, pp. 1--5.

\bibitem{Antminer_powerguide}
``{Zeus Mining Crypto Mining Pro: Antminer APW8 Power Supply Repair Guide [EN]},'' Published in 27, July 2010, [Online]. Available:~\url{https://www.zeusbtc.com/manuals/Antminer-APW8-Power-Supply-Repair-Guide.asp}.

\bibitem{9887746}
J.~Prakash and I.~Sarkar, ``{Comparison of PFC Converter Topology for Electric Vehicle Battery Charger Application},'' in \emph{2022 IEEE Students Conference on Engineering and Systems (SCES)}, 2022, pp. 1--6.

\bibitem{IEEEexample:PM_book}
P.~M. Anderson, B.~L. Agrawal, and J.~E. Van-Ness, \emph{Subsynchronous Resonance in Power Systems}.\hskip 1em plus 0.5em minus 0.4em\relax IEEE Power Engineering Society, 1989, pp. 7-14.

\bibitem{bollenUnderstandingPowerQuality2000}
M.~H.~J. Bollen, \emph{Understanding Power Quality Problems: Voltage Sags and Interruptions}, ser. {{IEEE Press}} Series on Power Engineering.\hskip 1em plus 0.5em minus 0.4em\relax {New York}: {IEEE Press}, 2000.

\bibitem{EMTP_PVPark}
U.~Karaagac, H.~Ashourian, I.~Kocar, A.~Stepanov, H.~Gras, and J.~Mahseredjian, ``{PV Park Models in EMTP},'' Published in 10, June 2021, Available:~\url{https://emtp.com/documents/EMTP%20Documentation/doc/devices-2022/Renewables/PV_Park_Models.pdf}, Tech. Rep., 2021.

\end{thebibliography}

\end{document}